# APPLICATION OF DEEP LEARNING FOR FACTOR TIMING IN ASSET MANAGEMENT

Prabhu Prasad Panda, Carnegie Mellon University, Pittsburgh, Pennsylvania, US
Maysam Khodayari Gharanchaei, Carnegie Mellon University, Pittsburgh, Pennsylvania, US
Xilin Chen, Carnegie Mellon University, Pittsburgh, Pennsylvania, US
Haoshu Lyu, Carnegie Mellon University, Pittsburgh, Pennsylvania, US

**ABSTRACT**

*The paper examines the performance of regression models (OLS linear regression, Ridge regression, Random Forest, and Fully-connected Neural Network) on the prediction of CMA (Conservative Minus Aggressive) factor premium and the performance of factor timing investment with them. Out-of-sample R-squared shows that more flexible models have better performance in explaining the variance in factor premium of the unseen period, and the back testing affirms that the factor timing based on more flexible models tends to over perform the ones with linear models. However, for flexible models like neural networks, the optimal weights based on their prediction tend to be unstable, which can lead to high transaction costs and market impacts. We verify that tilting down the rebalance frequency according to the historical optimal rebalancing scheme can help reduce the transaction costs.*

**Keywords:** Artificial Intelligence, Deep Learning, Factor Timing, Asset Management

## 1. INTRODUCTION

The practicality of timing factor investments remains a hotly debated topic in the financial world. A primary point of contention centres on the ability to predict factor premiums (Agrawal et al., 1992), which is crucial for successful factor timing (Black et al., 1972). However, (Asness, 2017) reveals that numerous traditional macroeconomic and fundamental variables lack significant explanatory power in a longer time frame and the timing is hard to implement. The current situation of factor timing research resembles the equity premium prediction (Basu, 1983) and (Bhandari, 1988) back in past, where adjustments to existing models and new models are introduced. Here we hope to see whether they will also help the exploration of factor timing.

## 2. DATA

We source our factor data from Fama and French (Breeden, 1979), accessible at Kenneth French's data library (Breeden et al., 1989) (monthly, January 1963 to December 2022), and our timing variable data from Welch and Goyal (2008), accessible at Goyal's website (Campbell and Thompson, 2008) (monthly, January 1871 to December 2022). We merge these 2 datasets, and the merged table comprises 714 months of values for factors and timing variables from July 1963 to December 2022 (around 60 years). We choose to time the Conservative Minus Aggressive (CMA) factor with Term spread (tms, which reflects the interest rate macro) and Default spread (Chan et al., 1985) (dfy, which reflects the prevailing credit condition) of the previous month from Welch and Goyal (2008), together its own 1-month lag (which handles the autocorrelation). As for training-testing split, we choose to use 1963-2002 as the training period (474 months) and 2003-2022 as the testing period (240 months). The data is of monthly frequency (more specifically, end-of-month).

| Variables | Descriptions | Source |
|---|---|---|
| Conservative Minus Aggressive (CMA) | Average return on the two conservative investment portfolios minus the average return on the two aggressive investment portfolios | Fama and French (1993) |
| Market excess return (Mkt-RF) | The excess return on the market, is the value-weighted return on all NYSE, AMEX, and NASDAQ stocks (from CRSP) minus the one-month Treasury bill rate (from Ibbotson Associates) | Fama and French (1993) |
| Term spread (tms) | Term spread (tms) is the difference between the long term yield (lty) on government bonds and the Treasury-bill (tbl) | Welch and Goyal (2008) |
| Default spread (dfy) | The Default Yield Spread (dfy) is the difference between BAA and AAA-rated corporate bond yields | Welch and Goyal (2008) |
| Corporate bond return (corpr) | Long-term corporate bond returns from Ibbotson's Stocks, Bonds, Bills and Inflation Yearbook | Welch and Goyal (2008) |

Table 1: dataset variables and descriptions

## 3. METHODS

### 3.1 Predictive Models for Factor Premium

To compute the factor-timing weights of the CMA factor, we need the conditional expected premium E[r$_{t+1}$] of the factor (Chan et al., 1985), we start with a predictor *f* that predicts the factor premium (Chan et al., 1985) with the features **x$_t$**:

$$r_{t+1} = f(X_t) + \epsilon_{t+1}$$

Referring to (Chen, 1991) and (Fama and French, 1993), we pick a series of candidate regression models from both linear and nonlinear regression models:

| Model | Special Configuration |
|---|---|
| Linear Regression - OLS | with restriction on coefficient and prediction in Campbell and Thompson (2008) |
| Linear Regression - Ridge | with penalty multiple of 1.0; without Campbell and Thompson (2008) restriction |
| Random Forest Regressor | with at most 6 leaves (Gu, Kelly and Xiu 2020) |
| Neural Network - NN3 | A 3-layer fully-connected neural network with 32, 16, and 8 neurons and ReLU activations for each layer (Gu, Kelly and Xiu 2020) |

Table 2: candidate return prediction models

To reflect the nature of prediction and model calibration, all models other than NN3 are fitted in an expanding window regression (Gu et al., 2020) (step = 1 month), starting from with the whole training period (Merton, 1973). NN3 is trained on the training only without expanding window to avoid overfitting (Merton, 1980) and handle computational complexity (Pettenuzzo et al., 2014). We also restrain the maximal number of leaf nodes for random forest regressor (Rapach et al., 2010) and keep the NN relatively shallow (Reinganum, 1981) to avoid overfitting, as these models are very flexible. The models will be evaluated in two dimensions: (1) the out-of-sample (OOS) R-squared as predictor of factor premium (as proposed by (Roll, 1983), (Rosenberg et al., 1985), and (Ross, 2013) we use OOS R-squared without demeaning, and (2) the OOS cumulative return (measured as the accumulated wealth with US$1 of initial investment).

### 3.2 Factor Timing

The research focuses on a two-asset portfolio between CMA factor and risk-free asset (Shanken, 1982) and (Shanken and Weinstein, 1990). We assume a risk averse (Sharpe, 1964) and (Welch and Goyal, 2008) utility function $E[r] - \gamma.\sigma_r^2$ (Shanken and Weinstein, 1990), and the optimal weight (Welch and Goyal, 2008) in the factor that can maximize the utility:

$$\omega_t^* = \frac{1}{\gamma} \cdot \frac{E[r_{t+1}]}{\sigma_r^2} = \frac{1}{\gamma} \cdot \frac{f(X_t)}{\sigma_r^2}$$

We use the 1-step-forward model prediction as conditional expected return and the expanding window variance as the denominator. As for the benchmark, we will use the unconditional (constant) optimal weight in the factor, which use the historical mean and variance from the training set.

## 4. RESULTS

### 4.1 In-sample coefficient significance

For OLS linear regression without Campbell and Thompson (2008) restrictions has a regression summary with statistical inference. However, due to the multicollinearity problem, the p-values tend to be inflated due to variance inflation. Under such circumstances, 1-step lag of CMA still shows statistical significance (p-value < 0.0005), indicating a significant serial correlation.

### 4.2 OOS R-squared in Factor Premium Prediction

We compare the OOS R-squared (with zero mean) of the candidate predictive models with expanding window regression (other than NN3). The outcome shows that more flexible structures like NN3 and random forest can better explain the variance in the factor premium of the unseen period.

| Model | OOS R-squared, zero-mean |
|---|---|
| Linear Regression - OLS (with Campbell and Thompson restrictions) | 0.024068 |
| Linear Regression - Ridge | 0.029150 |
| Random Forest Regressor | 0.035687 |
| Neural Network - NN3 | 0.102629 |

Table 3: OOS R-squared of each model

### 4.3 Performance of Factor-Timing Strategy (No Transaction Cost)

In evaluation of factor-timing strategies with different predictive model, we visualize the cumulative wealth paths for the investors following each strategy with US$1 of initial investment. The unconditional (constant) optimal weighting as the benchmark. To evaluate the performance of the model selected over different prevailing market conditions, we evaluated the models on 4 investment horizons separately: 1) the full testing period (2003-2022), 2) pre-Global Financial Crisis (GFC) period (2003-2007), 3) GFC and recovery period (2007-2015), and 4) post-Global Financial Crisis period (2015-2022). In this session, we assume no transaction cost. The wealth paths also the factor timing with NN3 outperforms the other strategies for the early phase of the full testing period, while its performance drops around the GFC, and bounce up around the COVID period, which coincide with the low-interest rate regime in early 2022. The underperformance of NN3 factor timing after GFC are likely due to the decay of explanatory power (as NN3 is not trained with expanding window data) as the testing data get temporally further and the potential regime switch(es), and the bottom out around COVID may be associated with the reversion to low interest rate in 2022, which defined a similar regime to the early 2000s that NN3 is trained until (which is consistent to the fact that its performance dropped after mid-2022, where fed funds rate started to hike). However, the optimal weights defined with NN3 obviously fluctuates much more fiercely than other models, which can lead to much larger erosion if transaction cost is considered. According to the wealth paths, factor timing with the random forest provides a stably better performance than both linear candidates and the unconditional optimal weighting, and it entails much lower downside risk than NN3. Among all strategies of factor timing, random forest often yields the highest Sharpe Ratio. Besides, it is noteworthy that the ridge (L2-regulated) linear regression provides very similar wealth path as the OLS linear regression under Campbell and Thompson restrictions.

However, factor timing with ridge will entail a slightly larger position changes than OLS, which can lead to larger erosion if transaction cost is considered.

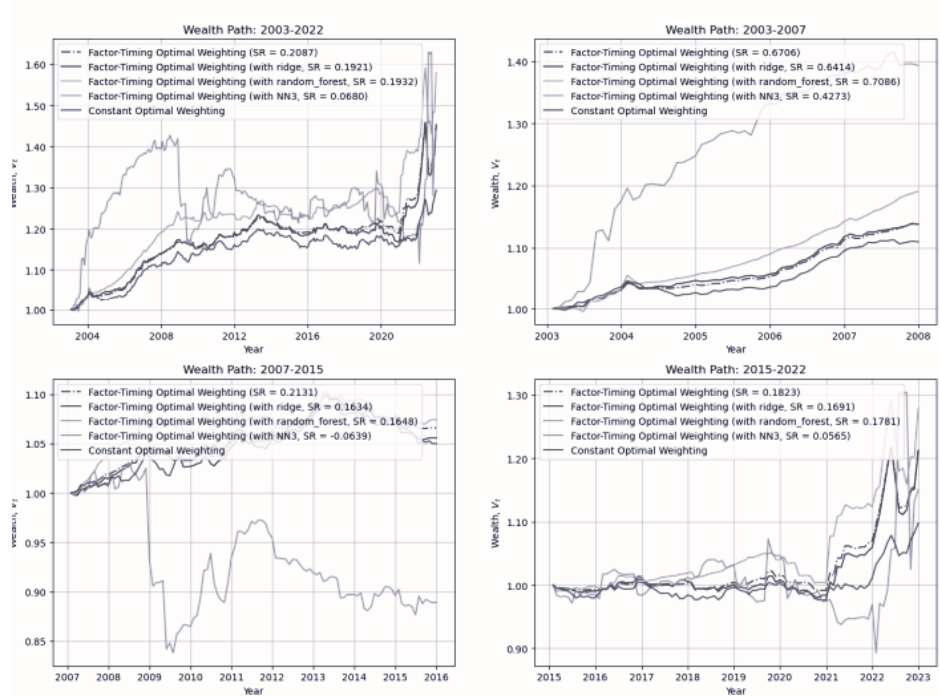

Fig. 1: Wealth paths of factor timing with different predictive models

| Model | 2003-2022 | 2003-2007 | 2007-2015 | 2015-2022 |
|---|---|---|---|---|
| Linear Regression - OLS (Campbell and Thompson) | 0.2087 | 0.6706 | 0.2131 | 0.1823 |
| Linear Regression - Ridge | 0.1921 | 0.6414 | 0.1634 | 0.1691 |
| Random Forest Regressor | 0.1932 | 0.7086 | 0.1648 | 0.1781 |
| Neural Network - NN3 | 0.0680 | 0.4273 | -0.0639 | 0.0565 |
| Constant (unconditional optimal) | 0.2141 | 0.4853 | 0.1232 | 0.1523 |

Table 4: Sharpe Ratio under no transaction cost assumptions

**4.4 Performance of Factor-Timing Strategy (with Proportional Transaction Cost)**

In this session, we assume that the transaction cost is proportional to the change in position. We observe that the more fluctuating weighting schemes, such as random forest NN3, will suffers larger return erosion than weighting schemes based on linear models. Starting from proportional transaction cost is 20bps, all models will underperform the constant weighting scheme. If the proportional transaction cost is 50bps, all models will underperform the constant weighting scheme. In this situation, we try to find the optimal rebalancing frequency to control the transaction cost erosion of the wealth. To avoid the problem with look-ahead bias, we pick the first 40% of the OOS months as the validation set and learn what is the optimal rebalancing interval (from 1 month to 12 months), and tilt down the rebalancing frequency of the latter 60% of the OOS months according to the optimal interval we observed on the validation set.

| | | | |
|---|---|---|---|
| Linear Regression - OLS (Campbell and Thompson) | 8 | 8 | 8 |
| Linear Regression - Ridge | 8 | 8 | 8 |
| Random Forest Regressor | 9 | 9 | 9 |
| Neural Network - NN3 | 1 | 11 | 11 |

Table 5: Optimal rebalancing interval (selected based on validation set) for proportional transaction cost

The outcome shows when the proportional transaction cost is low (e.g., 10bps), the difference between monthly rebalancing and valid-set-optimal rebalancing are trivial, while the difference gets wider as the fractional transaction costs get higher. For instance, if there exists a proportional transaction cost of 50bps, the optimal rebalancing interval for the factor timing with random forest and NN3 can lead to an annualized 0.13% and 0.63% of extra return on the latter 60% of the OOS months.

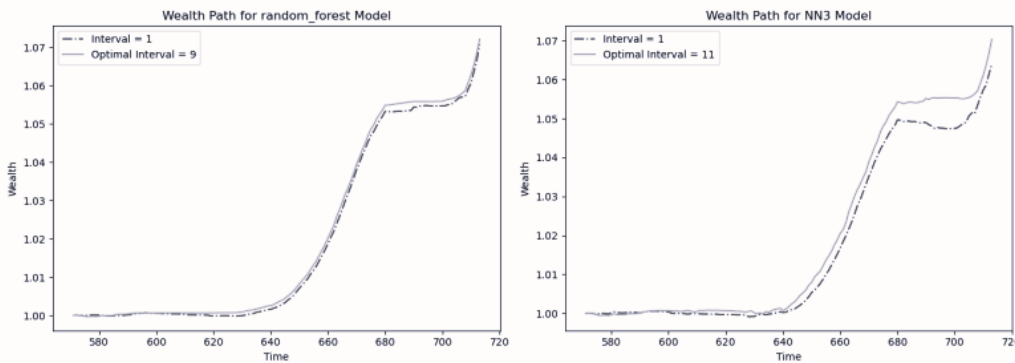

Fig. 2: Wealth improvement (US$1 invested started after the validation set) by using optimal rebalancing frequency, when proportional cost = 50bps

### 4.5 Discussion: Quadratic Transaction Cost

If we assume that the transaction cost is quadratic to the change in position, which is equivalent to linear market impact assumption. We observe greater impact on the wealth path on the longer term than on the shorter term, and the method we used under the proportional transaction cost does not work properly. If we have any opportunity, we hope to explore this topic further.

### 5. CONCLUSION

The project examines the performance of regression models on the prediction of CMA factor premiums. With the enhancement of Campbell and Thompson (2008), the OLS succeeds in guiding a factor timing strategy that outperforms the unconditional optimal weights. As for the other models, ridge linear regression provides very similar weights to the OLS with Campbell and Thompson (2008) restrictions, and the implementation is simpler, while we observe that the strategy with Ridge will have a higher turnover rate and entails higher transaction costs, which indicates that we cannot replace OLS with Campbell and Thompson (2008) restrictions with it. Besides, more complicated models like random forest and neural networks provide better explanatory power and factor timing performance, while they have an even higher turnover rate than linear models. Considering the transaction costs, 20 bps of proportional cost and 50 bps of quadratic cost will make all strategies with a US$ 1 initial investment underperforming the unconditional optimal weighting scheme. However, we need to be alert that expanding the initial investment by N times will lead to an N^2 times growth on the quadratic cost.

# 6. References and Bibliography